# Tunable Crystalline Order and Growth Kinetics in 2D Binary Colloidal Self-assembly Driven by Depletion Interactions

S.k. Tahmid Shahriar[1], Chris Feltman[2], Sean Machler[3], Nabila Tanjeem [3,*]

[1]Department of Mechanical Engineering, California State University, Fullerton, Fullerton, California, 92831, United States.
[2]Department of Mathematics, California State University, Fullerton, Fullerton, California, 92831, United States.
[3]Department of Physics, California State University, Fullerton, Fullerton, California, 92831, United States.

* Corresponding author: ntanjeem@fullerton.edu

**Abstract:** We investigate two-dimensional crystal assemblies formed by a binary mixture of colloidal particles with a size ratio of 0.88 and driven by short-ranged depletion interactions. Our experiments show that the orientational order of the assembly decreases with an increasing fraction of impurity particles, reaching up to 18% reduction in a 1:1 binary mixture compared to a monodisperse suspension. We observe slower growth rate and arrested dynamics in the binary mixture, whereas the monodisperse sample follows a two-step nucleation and growth mechanism. We performed molecular dynamics simulations to calculate the minimum energy states for different size ratios and number ratios of the binary mixture. The simulation results predict a compromised translational order but sustained orientational order in a binary mixture with a size ratio as high as 0.88. From the combined experimental and numerical results, we conclude that the disordered assemblies found in a binary mixture originate from the frustration in assembly kinetics rather than topological frustration when there is a marginal mismatch between the two particle sizes.

## 1. Introduction

Studies on colloidal self-assembly are instrumental in developing fundamental insights into material phases and designing applications by leveraging the mechanical and optical properties of the assembled phases.[1–5] The simplest components of colloidal self-assembly are monodisperse particles whose size, shape, interparticle interactions, and concentration determine the properties of the assembled phases. To realize greater complexity in the phase behavior and assembly kinetics, suspensions made of two different particle sizes, i.e., binary mixtures, have been employed in numerous experimental and computational studies.[6–13] Binary assemblies are the simplest model of polydisperse colloids that are the closest approximation to colloidal suspensions found in nature and relevant for many industrial applications. Insights gained from binary assembly experiments can be applied to understand the formation of metal alloys at the atomic scale[14,15] and to design materials with the desired crystalline structure.[16–20] Therefore, interest in understanding binary colloidal self-assembly for a wide range of experimental parameters, including particle size ratios, materials, compositions, spatial confinement, and interparticle interaction potentials, is rising rapidly.

An important determinant of colloidal self-assembly is the nature of the interparticle interaction potential. While hard sphere assembly has been a common choice for studying phase behavior, assembly mechanisms that involve engineering the attractive potential between particles by tuning electrostatic, DNA-hybridization, and depletion interactions have become increasingly popular.[21–23] Specifically, self-assembly driven by depletion forces allows for tuning the strength and the range of the attractive interaction potential simply by varying the assembly components, hence it has been employed for programming the structures of colloidal clusters.[24–28] Another important aspect of depletion interaction is that its strength depends on the geometry of the particles and the nearby surfaces; this property allows the formation of two-dimensional colloidal crystals on a surface whose curvature is smaller than the particle curvature.[29,30] Therefore, depletion-induced self-assembly has been an attractive choice for investigating 2D colloidal crystallization mechanisms on flat and curved surfaces.[31,32] Such systems have a diverse set of applications involving photonic crystals,[27,33] semiconducting particle assembly,[34] chiral nanomaterials,[31,35] photonic balls,[36,37], and colloidal epitaxy.[10,38] Despite significant interest, it is yet unknown how depletion-driven 2D colloidal crystallization is affected by particle polydispersity, the simplest case of which is a binary system comprised of two different particle sizes.

Here, we investigate two-dimensional binary colloidal self-assembly driven by a short-ranged depletion interaction potential, where the ratio of the small particle diameter to the large particle diameter is $\frac{d_1}{d_2} = 0.88$, and the range of the attractive interaction potential is only 4% of the particle diameter. Prior work on binary colloidal self-assembly in the presence of depletion interactions investigated the phase behavior in bulk,[39–41] observed eutectic crystallization,[42,43] and the formation of colloidal gels.[44,45] Toyotama et al. found phase-separated eutectic crystals for a size ratio of particles (0.83) that was closer to our system but driven by a long-ranged interaction potential (80% of particle diameter). Pandey et al. investigated self-assembly with short-ranged depletion interaction (4% of particle diameter) in a binary mixture of particles with a size ratio of 0.49 and observed the formation of 3D colloidal gels and arrested particle dynamics.

To our knowledge, no experiments have explained the formation of 2D binary crystals driven by short-ranged depletion interactions, especially in the presence of a small asymmetry in the particle sizes. Therefore, we performed a combination of experimental and computational investigation to elucidate how the size ratio and the relative concentrations of the two particle species determine the structural order and the assembly kinetics.

We found that the orientational hexatic order parameter of the self-assembled binary monolayer reduces by 18% in a 1:1 binary mixture compared to a monodisperse suspension. Additionally, the binary mixture exhibits much slower growth kinetics that is accompanied by cooperative particle motion. To explain the experimental findings, we developed a molecular dynamics simulation model to identify the minimum energy states of a 2D binary assembly driven by a short-range attractive interaction potential. The model shows that for a mixture with a greater particle size difference, i.e., $\frac{d_1}{d_2} \leq 0.86$, the hexatic order parameter decreases with an increasing fraction of impurity particles and reaches the minimum value for the 1:1 mixture. However, for a binary mixture where the particle sizes are much closer, i.e., $\frac{d_1}{d_2} = 0.88$, the orientational hexatic order parameter in the 1:1 mixture ratio sample remains the same as the sample with minimal impurity. However, in this case, a reduction in the translational order, as characterized by the average bond length, is observed. The disagreement between the computational and experimental results can be attributed to the assembly kinetics that distinguish the 1:1 mixture from a monodisperse sample. Although both samples grow in amorphous states during the first part of the assembly process, the monodisperse sample starts forming stable crystallites following two-step nucleation,[30] whereas the binary sample continues to grow in the amorphous state until the assembly attains high surface coverage and the particles become dynamically arrested. Our results provide a new insight into binary assemblies in cases where the mismatch in the particle sizes is marginal — the frustration in assembly kinetics, rather than the topological frustration, is the key factor that determines the structural order of the assembly.

## 2. Materials and Methods

### 2.1 Preparation of colloidal particle mixtures:

Fluorescent polymer microspheres with two different diameters, $d_1 = 700$ nm (Fluoro-Max G700, green beads purchased from Fisher Scientific Inc.) and $d_2 = 790$ nm (Fluoro-Max R800, red beads purchased from Fisher Scientific Inc.) were used to prepare the binary mixture samples. Each of the particle suspensions was washed two times before usage. During each wash cycle, a total of 1 mL particle suspension (1% solids) was centrifuged (using Eppendorf 5430) for 5 minutes at 6500 rpm, the supernatant was removed, and the particles were re-suspended in 1 mL of Milli-Q water. The particles were redispersed in the suspension through vortex-mixing for 15-20 s. Once the microsphere suspensions were prepared, they were mixed with an aqueous 36.5 mM of sodium dodecyl sulfate (SDS, purchased from Sigma Aldrich) solution for the depletion-induced self-assembly experiments. A stock solution of 100 mM SDS was first prepared by mixing 288.38 mg of SDS in 10 mL of Milli-Q water. Then the SDS solution was diluted to 73 mM by mixing 36.9 µL Milli-Q water with 100 µL of the stock SDS solution. Binary mixtures with different volume ratios were prepared as described below:

Monodispserse sample: The green beads (Fluoro- Max G700) with a diameter of $d_1 = 700$ nm were used to study the self-assembly of a monodisperse suspension. Five µL of the 1% particle suspension was mixed with an equal volume of 73 mM of SDS. 1:1 binary mixture: 10 µL of 73 mM SDS solution was mixed with 5 µL of 700 nm green beads and 5 µL of 790 nm red beads. 9:1 binary mixture: 9 µL of 700 nm green beads and 1 µL of 790 nm red beads were mixed with 10 µL of 73 mM SDS solution. 19:1 binary mixture: 19 µL of 700 nm green beads and 1 µL of 790 nm red beads were mixed with 20 µL of 73 mM SDS solution. 39:1 binary mixture: 39 µL of 700 nm green beads and 1 µL of 790 nm red beads were mixed with 40 µL of 73 mM SDS solution.

**2.2 Sample chamber preparation and imaging:**

After the colloidal particles were mixed with SDS, the solution was transferred to a sample chamber for observation under a microscope. A 2-D sample cell was prepared by making a sandwich of a 24x50mm cover glass (purchased from VWR, no. 1) and a 22x22mm cover glass (purchased from VWR, no. 1). First, both cover glasses were pretreated in a UVO Cleaner (Jelight Model 30) for 5 minutes. A thin plastic film spacer and a 22x22mm micro cover glass were placed on top of the 24x50mm cover glass, and the corners of the chamber were sealed by applying the optical adhesive NOA 61 (Norland). The sample cell was then re-treated in the UVO Cleaner again for a minute. The particle suspension (10 µL) was injected into the thin gap between the two cover glasses. Finally, the sample cell was completely sealed by applying a 5-minute epoxy (Devcon).

The sample cells were imaged using an optical microscope immediately after sealing them. The imaging was performed using an inverted Olympus ix83 microscope equipped with a 100x (N.A. = 1.45) objective lens. Images were captured using the open-source software Fiji and its plug-in Micro-Manager 2.0. The green particles were captured using a 470 nm/525 nm (excitation/emission) filter set, and the red particles were captured using a 560 nm/ 630 nm filter set.

## 2.3 Image processing and analysis:

**Identification of particle positions:** The images of the colloidal crystals were first converted to 8-bit images using Fiji. An open-source Python package - Trackpy, was used to identify the particles and locate their positions from the image files. The particles were identified by optimizing four different parameters – particle diameter, minimum separation distance between the neighboring particles, minimum brightness of the features, and the overall brightness profile. The estimated particle diameter was calculated from the microscopic images using Fiji, and the minimum separation distance was kept at a value equal or close to the particle diameter.

**Analysis of colloidal crystal images:** Three different quantitative analysis techniques were performed to characterize the features of the colloidal crystals - Voronoi diagram, hexatic order parameter, and radial distribution function. An in-house Python script was developed to calculate the Voronoi diagrams and the average hexatic order parameter. The positions (x,y coordinates) of the particles were identified using Trackpy, and an existing Python library (Voronoi from SciPy) was used to generate the Voronoi diagrams. Another Python library (NearestNeighbors) was used to identify the coordinates of the six nearest neighbors of every particle. A cut-off distance was applied to exclude particles that are greater than $1.5D - 2D$ (D: particle diameter) away from the nearest neighbor list. Then, the hexatic order parameter of each particle was calculated from the following equation: $\psi_6(k) = \frac{1}{n}\sum_{j=1}^{n} e^{i6\theta_{jk}}$, where $\theta_{jk}$ is the angle between the vector **r**$_{jk}$ that connects particle *k* to particle *j*, and the x-axis. An existing Python library (PairCorrelation) was used to calculate the radial distribution function g(r) from colloidal crystal images.

**Analysis of particle dynamics:** Particles were tracked from a series of images to quantify assembly dynamics over time. First, particle locations were identified from all relevant image frames using Trackpy, and then their positions were linked to calculate the trajectories of each particle. The process was repeated for both the red and the green particles, images of whom were acquired using different fluorescent channels. Quiver plots were calculated to visualize the cooperative motion of the particles from the magnitude and the direction of their displacement vectors. To achieve this plot, the displacement vectors were calculated from the initial position and the final position of the particles identified from their trajectories.

## 2.4 Molecular dynamics simulation:

**Defining interaction potentials:** The Lennard-Jones interaction potential was employed to emulate particles with different finite sizes and short-ranged interactions. Interaction between type 1 (small) particles:

$$U_1(r) = 4\varepsilon_1 \left[\left(\frac{\sigma}{r-\Delta_1}\right)^{12} - \left(\frac{\sigma}{r-\Delta_1}\right)^6\right], \quad r < r_C + \Delta_1 \quad (1)$$

Interaction between type 2 (large) particles:

$$U_2(r) = 4\varepsilon_2 \left[\left(\frac{\sigma}{r-\Delta_2}\right)^{12} - \left(\frac{\sigma}{r-\Delta_2}\right)^6\right], \quad r < r_C + \Delta_2 \quad (2)$$

Interaction between type 1 (small) and type 2 (large) particles:

$$U_{12}(r) = 4\bar{\varepsilon} \left[\left(\frac{\sigma}{r-\bar{\Delta}}\right)^{12} - \left(\frac{\sigma}{r-\bar{\Delta}}\right)^6\right], \quad r < r_C + \bar{\Delta} \quad (3)$$

The finite diameters the smaller and the larger particles were represented by $\Delta_1 + \sigma$ and $\Delta_2 + \sigma$, respectively. Five different size ratios $m = 0.80, 0.82, 0.84, 0.86, 0.88$ were applied to simulate different degrees of asymmetry in particle sizes. The values of $\Delta_2$ and $\bar{\Delta}$ were calculated for each $m$ and for a fixed value of $\Delta_1$ (all parameters listed in Table 1). The value of $\sigma$ was calculated from the experimental estimate of the range of depletion interaction - 4% of the diameter of the smaller particles, i.e., $\frac{\sigma}{\Delta_1+\sigma} = 0.04$. Since the ratio of the interaction strengths, $\frac{\varepsilon_1}{\varepsilon_1} = m$ is true for depletion forces, $\varepsilon_2$ and $\bar{\varepsilon}$ were calculated from the values of $\varepsilon_1$ and $m$. The cutoff distance $r_c$ was chosen to be approximately $1.5\,[(\Delta_1 + \sigma) + (\Delta_2 + \sigma)]$ to reduce the computation time.

**Table 1:** LJ potential parameters used in simulations

| Size ratio $m = \frac{\Delta_1 + \sigma}{\Delta_2 + \sigma}$ | $\Delta_1$ | $\Delta_2$ | $\bar{\Delta} = \frac{\Delta_1 + \Delta_2}{2}$ | $\sigma$ | $\frac{\sigma}{\Delta_1 + \sigma}$ | $\varepsilon_1$ | $\varepsilon_2$ | $\bar{\varepsilon} = \frac{\varepsilon_1 + \varepsilon_2}{2}$ |
|---|---|---|---|---|---|---|---|---|
| 0.8 | 4.45 | 5.61 | 5.03 | 0.19 | 0.04 | 5 | 6.25 | 5.63 |
| 0.82 | 4.45 | 5.47 | 4.96 | 0.19 | 0.04 | 5 | 6.1 | 5.55 |
| 0.84 | 4.45 | 5.33 | 4.89 | 0.19 | 0.04 | 5 | 5.95 | 5.48 |
| 0.86 | 4.45 | 5.19 | 4.82 | 0.19 | 0.04 | 5 | 5.81 | 5.41 |
| 0.88 | 4.45 | 5.08 | 4.77 | 0.19 | 0.04 | 5 | 5.68 | 5.34 |

**Initialization of the simulation region:** The simulations were performed in a box with dimensions 800 x 800 x 1, and particle dynamics were limited in only x and y dimensions. $N_1$ number of type 1 particles (diameter $d_1 = \Delta_1 + \sigma$) and $N_2$ number of type 2 particles (diameter $d_2 = \Delta_2 + \sigma$) were placed in random positions of the box at the beginning of the simulation. The values of $N_1$ and $N_2$ were chosen carefully to maintain a constant total surface coverage in all simulations performed for different $\frac{N_1}{N_2}$ cases. The following equations were solved to calculate $N_1$ and $N_2$.

$$N_1 \pi \left(\frac{d_1}{2}\right)^2 + N_1 \pi \left(\frac{d_1}{2}\right)^2 = A \quad (4)$$

where $A \approx 425{,}000$ is the total area covered by all particles in the simulation box, and

$$\frac{N_1}{N_2} = \frac{1}{90}, \frac{1}{9}, 1 \qquad (5)$$

are the three different number ratios that were applied in the simulations. The determined values of $N_1$ and $N_2$ are summarized in Table 2.

**Table 2:** The number of different particle types in the simulation box

|  | $N_1 : N_2$ | $N_1$ | $N_2$ |
|---|---|---|---|
| $m = 0.80$ | 90: 1 | 24713 | 273 |
|  | 9:1 | 21420 | 2381 |
|  | 1:1 | 9808 | 9827 |
| $m = 0.82$ | 90: 1 | 24727 | 272 |
|  | 9:1 | 21572 | 2353 |
|  | 1:1 | 10102 | 10077 |
| $m = 0.84$ | 90: 1 | 24727 | 272 |
|  | 9:1 | 21757 | 2449 |
|  | 1:1 | 10836 | 10840 |
| $m = 0.86$ | 90: 1 | 24732 | 273 |
|  | 9:1 | 21769 | 2435 |
|  | 1:1 | 10607 | 10639 |
| $m = 0.88$ | 90: 1 | 24727 | 283 |
|  | 9:1 | 21947 | 2459 |
|  | 1:1 | 10976 | 10984 |

**Simulated annealing:** We applied Langevin dynamics to reduce the temperature of the simulation gradually from $T_i = 5\varepsilon/k_B$ to $T_f = 0.1\varepsilon/k_B$ over a period of 500000 iterations, with the time step of each iteration, τ = 0.001 s. We confirmed that closely packed crystalline structures were found at the end of the simulation period for every set of parameters.

## 3. Results and Discussions

We formed 2D binary colloidal assemblies on a glass surface by preparing a ternary mixture of polystyrene spheres with two different diameters, $d_1$ = 700 nm and $d_2$ = 790 nm, and 36.5 mM of sodium dodecyl sulfate (SDS) (Figure 1a). The polydispersity index of the polystyrene particles was less than 3%, as reported by the supplier. The SDS surfactants form micelles at the applied concentration and induce attractive depletion interactions between the polystyrene spheres and the flat glass surface, as well as between the polystyrene spheres themselves.[31,46] The Asakura-Osawa depletion interaction potential between two objects can be expressed as:[47]

$$U_{AO}(r) = -\rho k_B T V_{ov}(r)$$

where $\rho$ is the number density of the depletant micelles and $V_{ov}$ is the overlap of the excluded volumes, as illustrated in Figure 1b. Because the overlap volume between a flat surface and a sphere is greater compared to the overlap volume between two spheres, the particles first weakly adsorb to the flat glass wall at the bottom of the sample chamber and then form closely packed structures (Figure 1b). The SDS concentration was carefully chosen to keep the strength of the attractive potential within a few $k_B T$ so that equilibrium assembly can take place through multiple particle binding and unbinding events.

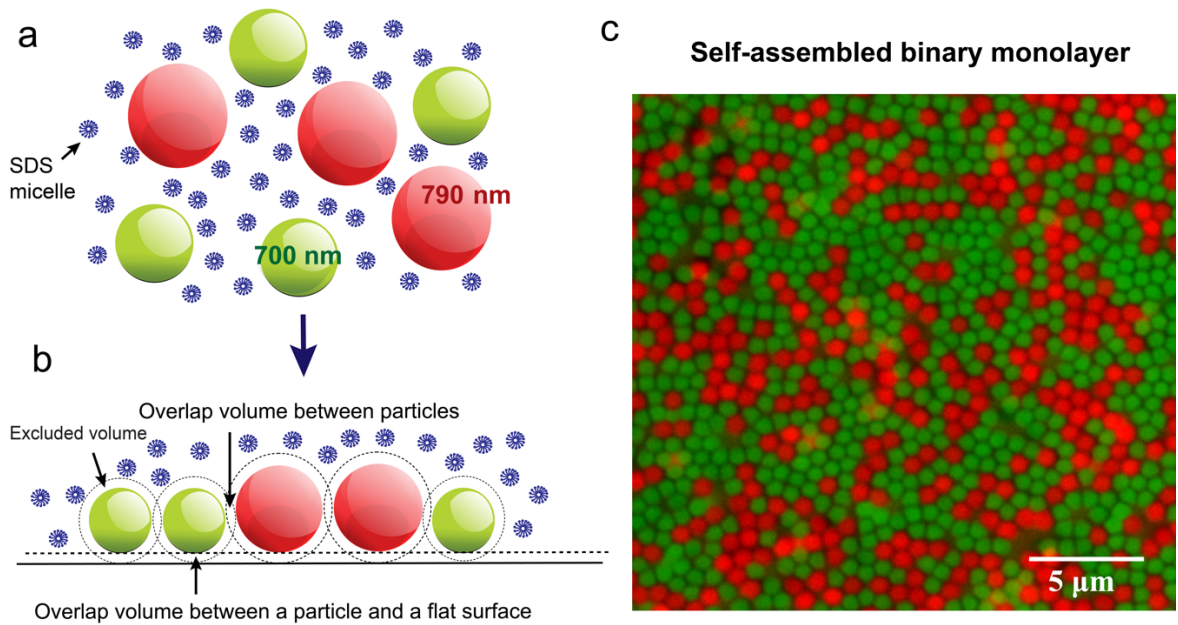

Figure 1 (a) An illustration of the ternary mixture – the binary mixture colloidal polystyrene spheres (green: 700 nm and red: 790 nm) and the SDS micelles. (b) An illustration of the assembly mechanism driven by depletion interactions. The overlaps of the excluded volumes are shown for the particle-particle geometry and particle-flat surface geometry. A closely packed monolayer on the flat surface forms when the maximum overlap volume is achieved minimizing the free energy of the system. (c) Optical micrograph of a self-assembled layer formed by a binary mixture of 1:1 volume ratio.

The structure and dynamics of the self-assembled monolayers were observed in real-time using an optical microscope (Olympus ix83) equipped with fluorescence imaging capabilities and a 100x oil immersion objective lens (NA = 1.45). Particles with diameter $d_1 = 700$ nm and $d_2 = 790$ nm were identified from their fluorophore labels - green and red colors, respectively. The self-assembled monolayers formed after leaving the mixture solution in a sealed sample chamber for about an hour (Figure 1c).

First, we investigated how the mixing ratio of the two types of particles impact the crystalline order of the formed layers. We calculated the hexatic order parameter that describes the degree of six-fold rotational symmetry around every particle, $\psi_6(k) = \frac{1}{n}\sum_{j=1}^{n} e^{i6\theta_{jk}}$, to quantify the crystalline order. In a sample containing only one of the particles ($d_1$ = 700 nm), polycrystalline structures were observed with an average hexatic order of $\overline{\psi_6} = 0.74$. The average $\overline{\psi_6}$ was calculated from micrographs of ten different positions in a sample, each consisting of at least 550 particles. Next, we mixed the 700 nm particles (green) with 790 nm particles (red) in a few different volume ratios – 39:1, 19:1, 9:1, and 1:1 (Figure 2). As the fraction of "impurity" particles, i.e., 790 nm (red) particles, increases in the mixture, the hexatic order gradually decreases, as shown in Figure 2. The minimum value of $\overline{\psi_6}$ was found to be at 0.61 for a 1:1 mixture sample after 90 min of observation. This result indicates that the self-assembled structures transition from crystalline to amorphous for a higher fraction of impurities and thereby offers a mechanism for precise tuning of the crystalline order in 2D colloidal assemblies.

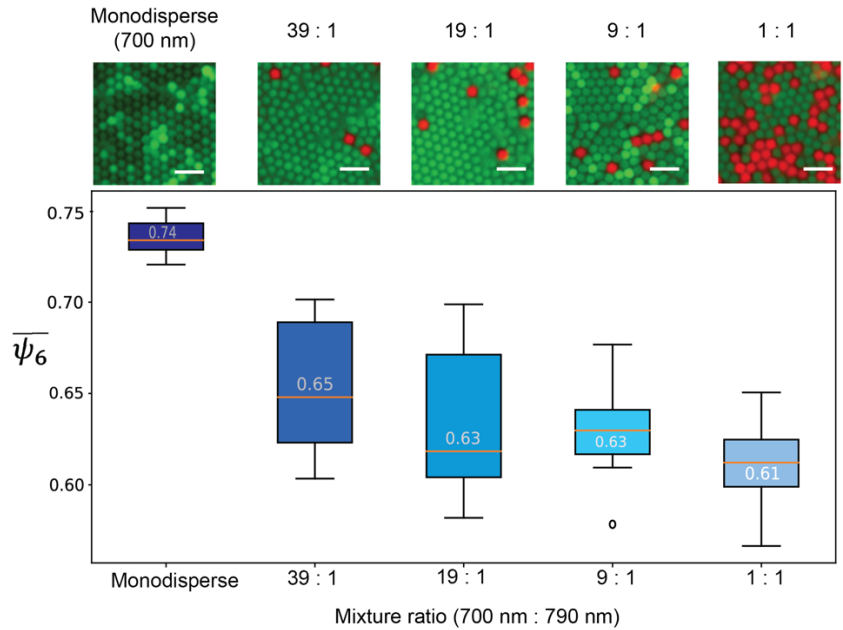

Figure 2 The average hexatic order decreases with the increase in the fraction of the impurity (red) particles. Each box plot represents the distribution of the hexatic order parameter for each different mixture ratios – homogenous (only green particles), 39:1, 19:1, 9:1, and 1:1, calculated from different positions of the sample chamber. The median hexatic order $\overline{\psi_6}$ is marked by the orange horizontal line on the respective boxes. The five images above show representative micrographs for the corresponding mixture ratios (scale bar 3 μm).

To understand the origin of the tunable crystalline order in depth, we developed a molecular dynamics simulation model that allows us to identify the equilibrium assembly structures for a wider range of parameters. The simulation was performed using the open-source molecular dynamics simulation package LAMMPS.[48] The different sizes of the particles were implemented by using shifted Lennard-Jones (LJ) potentials (equations 1-3), where $\Delta_1 + \sigma$ represents the diameter of the smaller particles (green) and and $\Delta_2 + \sigma$ represents the diameter of the larger particles (red).[49] The range of the attractive interaction potential was fixed at 4% of the smaller particle diameter to mimic the effective diameter of the depletant micelles (about 30 nm) estimated in prior studies.[46,50] The shapes of the applied LJ potential between different particle pairs are shown in Figure 3a.

We found the minimum energy states of the 2D assembly for different size ratios, $m = \frac{d_1}{d_2} = 0.80, 0.82, 0.84, 0.86, 0.88$ by performing simulated annealing. We observe a reduction in the average hexatic order, $\overline{\psi_6}$, with increasing fractions of impurities, specifically for cases where the difference in the particle sizes was larger (smaller $m$). For example, $\overline{\psi_6}$ decreases from 0.89 to 0.69 as the mixture ratio varies from 90:1 to 1:1 when the simulation is performed at $m = 0.80$ (Figure 3b) However, the average hexatic order $\overline{\psi_6}$ does not decrease when two particle types have closely matched diameters, i.e., $m = 0.88$ – a size ratio that represents the experimental sample. The assembled structures found at the minimum energy states are shown in Figure 3c - polycrystalline structures form when the fraction of impurities is small (90:1) for both $m = 0.80$ and $m = 0.88$. In contrast, evenly distributed defects and voids are observed in the 50% mixture ratio (1:1) sample (Voronoi diagrams in Figure S1). The constant $\overline{\psi_6}$ value in the 1:1 and the 90:1 mixture for $m = 0.88$ indicates that the energy cost from the grain boundaries in the 90:1 mixture matches the energy cost of the voids in the 1:1 mixture. As the mismatch in particle diameters becomes larger, e.g., for $m = 0.80$, the number of defects increases, causing a significant reduction (22%) in the average hexatic order. In summary, the computational results indicate that the formation of amorphous solids in the presence of a large size mismatch can be explained primarily from the energetics of the system. This finding is consistent with prior experiments that observed glass formation in binary samples with a size ratio of 0.77 and 0.81.[51,52]

Additionally, we carefully examined the translational order of the assembled structures by quantifying the distributions of the center-to-center distance between neighboring particles, $D_B$. As shown in Figures 3d and 3e, the average $\overline{D_B}$ and the standard deviation of the distribution, $\delta D_B$ gradually increases with the increasing fraction of impurities (both for $= 0.80$ and $m = 0.88$). This increase is more prominent in the presence of large size mismatch as demonstrated in Figure 3f. The finding indicates that the translational order is more sensitive to marginal mismatches in particle diameters than the orientational hexatic order. Although the asymmetric bond lengths between particle pairs reduce the average translational order, the deviation is not high enough to compromise the mean orientational order in the 1:1 binary mixture for $m = 0.88$.

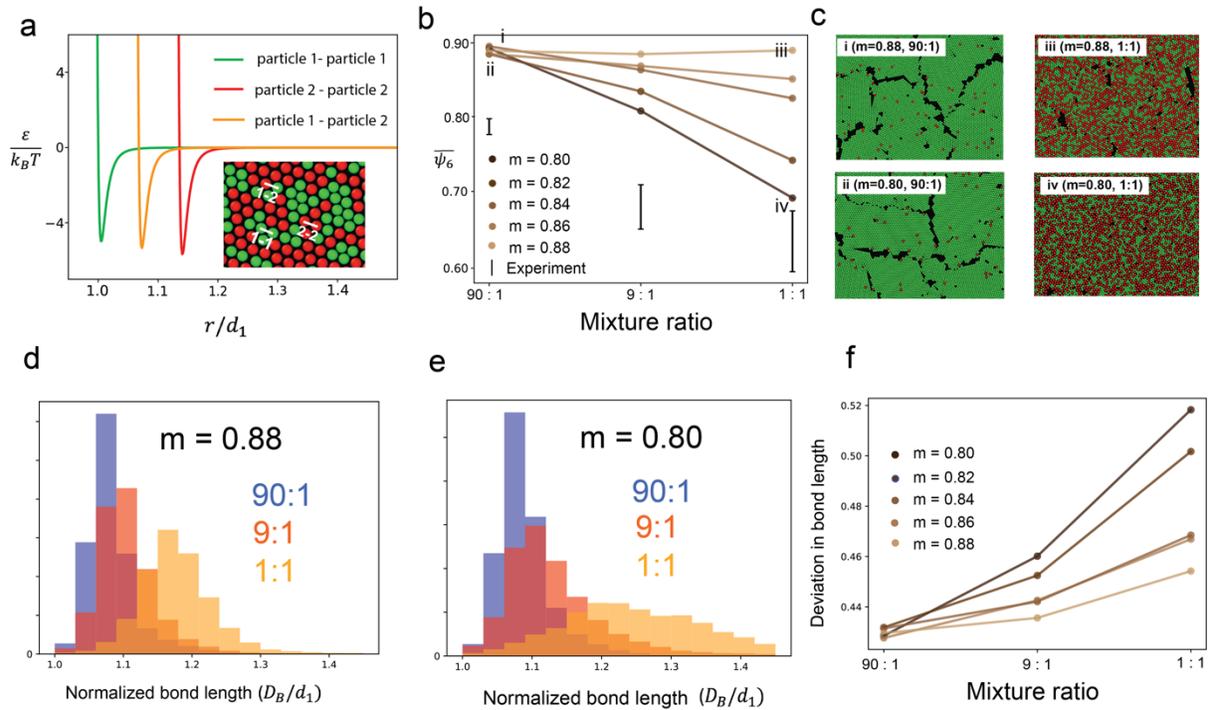

Figure 3 Simulation results (a) The Lennard-Jones interaction potentials between different particle pairs: the green line shows the interactions between the smaller particles (particle 1), the red line shows the interactions between the larger particles (particle 2), and the orange line shows the interactions between the smaller and the larger particles. (b) The average hexatic order $\overline{\psi_6}$ of the minimum energy state structures for different size ratios and number ratios, showing that the simulated orientational order decreases rapidly for the m$= 0.80$ sample but remains constant for the $= 0.88$ sample. The experimental data points were taken for m $= 0.88$. (c) The images of colloidal crystals at the ground state for four different parameter sets depicted by the data points annotated as (i), (ii), (iii) and (iv) on the plot of Figure 3b. (d) The distribution of bond lengths for m $= 0.88$ showing that the mean bond length increases with increasing fraction of impurities. (e) The distribution of bond lengths for m $= 0.80$ shows a similar trend but with a much wider distribution for the 1:1 mixture ratio. (f) The standard deviation of the bond lengths calculated from the distributions of all samples. Although smaller compared to other cases, the m $= 0.88$ sample shows a wider distribution of bond lengths as the number ratio varies from 90:1 to 1:1.

We examined the hexatic order for the $m = 0.88$ sample numerically for an even shorter range of attractive interaction – 0.22% of the particle diameter. We found that the hexatic order shows only a 2% reduction when the number ratio varies from 90:1 to 1:1, much smaller compared to the 18% reduction observed in the experiment. Furthermore, to investigate the thermodynamic equilibrium states, we excited the ground state crystal structures with an increased temperature of $k_B T$. We found that the hexatic orders of all mixture ratios (90:1, 9:1, and 1:1) decrease equally from 0.89 to 0.82 at room temperature and any notable reduction was absent in the 1:1 sample compared to the 90:1 sample (Figure S2). Therefore, we conclude that the near equilibrium structures computed in the simulations do not explain why we observe a large reduction of $\overline{\psi_6}$ (18%) in the $m = 0.88$ sample when the mixture ratio changes from 90:1 to 1:1.

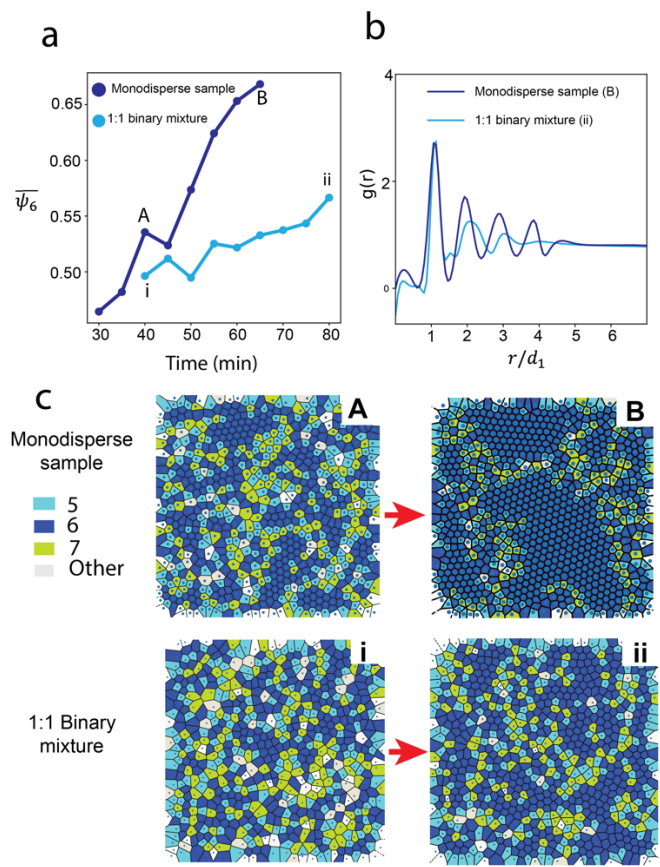

Figure 4 Comparison of crystal growth kinetics between a monodisperse sample and a 1:1 binary mixture sample. (a) The time evolution of the average hexatic order $\overline{\psi_6}$ shows a significantly slower ordering mechanism for the binary mixture compared to the monodisperse sample. (b) The radial distribution function obtained from the structures found at the end of the measurement - the monodisperse sample at 65 min and the binary mixture at 80 min. (c) The Voronoi diagrams of the two samples at different time points illustrating the crystallization dynamics. The monodisperse sample forms stable crystallites in between the timeframe of 40 min (A) – 65 min (B). In contrast, the binary mixture shows amorphous structures both at 40 min (i) and at 80 min (ii), with a small increase in the crystalline order over a long period of time.

Therefore, we shift our focus to the investigation of the crystal growth kinetics with the goal of explaining the experimental observations. To obtain a quantitative description of the growth, we captured the bond network and calculated the $\overline{\psi_6}$ value every 5 min intervals for both the monodisperse sample and the 1:1 binary mixture sample. As demonstrated in Figures 4a and 4c, the particles remain in an amorphous state in both samples with a low $\overline{\psi_6}$ in the range of 0.50 – 0.55 (point A and point i) for the first half an hour. However, the growth kinetics show major difference starting from 40 min when the monodisperse sample rapidly nucleates into a solid state to form two stable crystalline grains (point B at 60 min). During the same timeframe, the 1:1 sample remains in an amorphous state with a low $\overline{\psi_6}$ that only reaches a value of 0.56 after 80 min. The Voronoi diagram of the 1:1 sample at this time (point ii) contains small and unstable crystalline regions surrounded by many 5- and 7-fold defects. The short-range order in the 1:1 sample is evident from the radial distribution function plots of Figure 4b – which exhibits reduced heights of the higher-order peaks compared to the monodisperse, crystalline sample.

Finally, we captured the microscopic dynamics of the amorphous state in the 1:1 binary mixture sample after 80 min (Movie S1). We observed multiple partial vacancies in the assembly and cooperative particle movement adjacent to the locations of the partial vacancies. An example of the cooperative motion is presented and quantified in Figure 5. Because of the remarkably short interaction range, particles fail to interact with each other across a partial vacancy and rearrange their positions locally due to their Brownian motions. These local movements result in continuous hopping of the vacancy locations (Figure 5a). Because of the high surface coverage, the fluctuating particles can support the relocation of their nearest neighbors, causing the cooperative movement observed in certain spots (Figure 5b). Particle motion of this nature contributes to the overall arrested dynamics, which is a common feature observed near glass transition.[53,54]

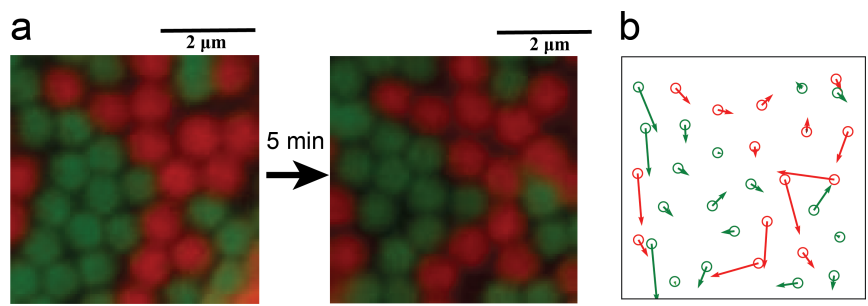

Figure 5 Microscopic particle dynamics in an amorphous 1:1 binary mixture sample. (a) The snapshots of a self-assembled region within 5 min time interval show minute rearrangement of particle and vacancy positions. (b) A quiver plot quantifying and visualizing the displacements of all particles from (a) during the examined time interval. The cooperative motion of at least three groups of particles is observed as depicted by the groups of arrows that point in the same direction.

Colloidal crystallization driven by short-ranged depletion interaction is known to follow a two-step nucleation pathway, distinct from the mechanism explained in the classical nucleation theory.[30] Savage et al. found that especially at a high area fraction, the assembly first forms amorphous clusters and grows in that state until a sudden increase in the hexatic order occurs resulting a crystalline state. The mechanism behind this two-step nucleation was explained by the reduction of the line tension in the intermediate amorphous phase that compensates for the lower bulk energy caused by the short interaction range. This is indeed what we observe in the crystallization pathway of the monodisperse sample (Figure S3) – the assembly first grows in the amorphous state, and then stable crystallites start developing after 40 min (Figure 4a and S3), which eventually grow and create boundaries between multiple crystalline grains. However, in the binary sample, when a marginal mismatch in the particle size and interaction potential is introduced, the growth in this amorphous state continues for the entire timespan of the assembly, as depicted by the slow growth curve in Figure 4a and snapshots in Figure S3. The prolonged stage of amorphous growth is likely due to the size mismatch that causes a further reduction of the bulk energy in a binary simple compared to a monodisperse sample – as shown in the higher nucleation barrier in binary systems.[55] The mobility of the particles gradually decreases with the absorption of new particles into the monolayer, giving rise to the arrested dynamics and a compromised crystalline order. Furthermore, when a stronger interaction potential is applied by increasing the concentration of the depletant micelles, the binary sample forms colloidal gels (Figure S4), transitioning from one nonequilibrium phase to another.

## 4. Conclusions

We investigated the self-assembly of 2D binary colloidal crystals driven by short-ranged depletion interaction and in the presence of a marginal mismatch in the particle diameters. Our computational findings indicate that the minimum energy states maintain a high orientational order in a binary sample when the size ratio of the small to the large particle is 0.88. However, our experiments show a significant reduction (from 0.74 to 0.61) of the orientational order parameter in the 1:1 binary mixture compared to the monodisperse suspension. The order reduces gradually as the fraction of the impurity particles increases in the binary mixture. By carefully examining the crystallization kinetics, we find that the two-step nucleation mechanism that drives crystallization in the presence of short-ranged depletion interaction is responsible for the observed phenomenon. After growing in the amorphous state during the first step, the monodisperse sample enters the second nucleation step to rapidly form stable crystallites while the binary sample continues to grow in the amorphous state until the particle dynamics becomes arrested. These results have important implications for our general understanding of the mechanisms behind amorphous solid formation that originate from thermodynamic, kinetic, or topological conditions. Our findings on 2D binary assembly demonstrate that the frustration in growth kinetics can dominate over the topological frustration when a small mismatch in particle size is introduced. Leveraging this kinetic pathway of colloidal glass formation can stimulate progress in the fundamental studies on glass transitions [54,56–58] as well as applications focused on photonic glasses.[59–61] In addition, the binary mixtures with intermediate volume ratios (19:1 and 9:1) allow the nucleation of small crystallites whose growth is then inhibited by the presence of the impurities. Applications of this self-limiting formation of crystal grains could enable the design of polycrystalline and epitaxial colloidal materials [10,62–65] with precisely tuned optical and mechanical properties.

# Appendix

1. **Analysis of the simulated crystalline structures**

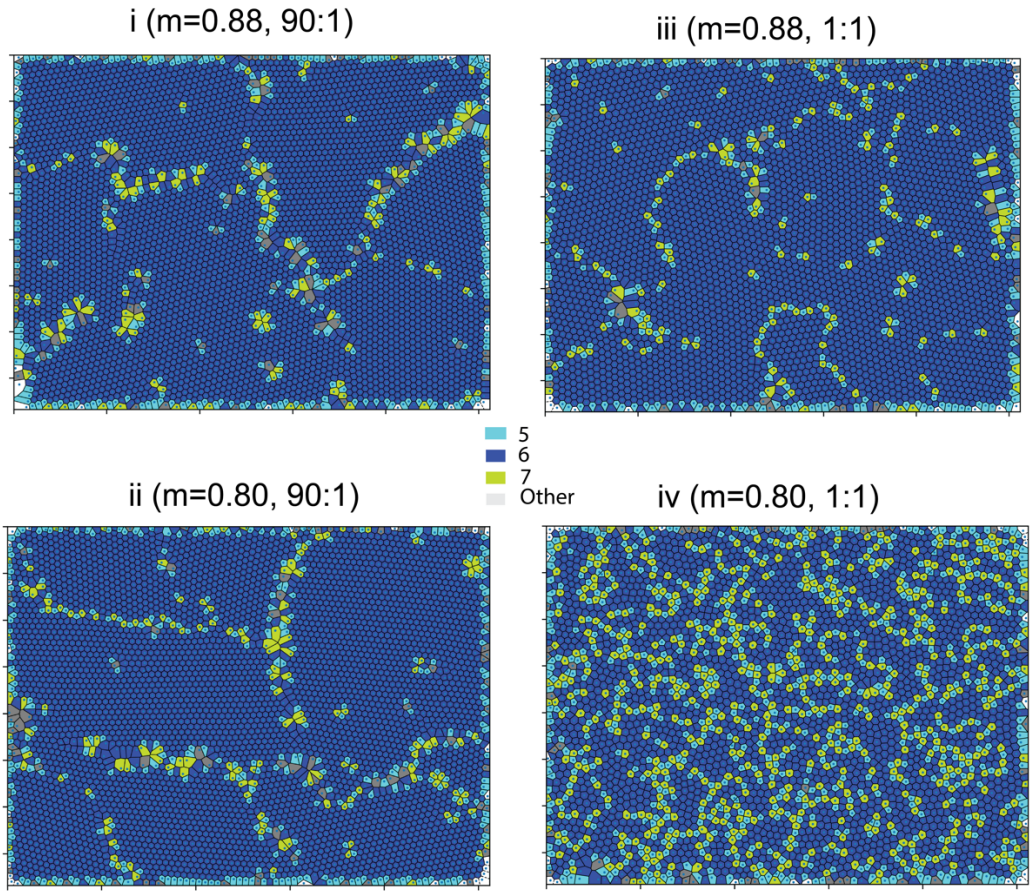

Figure S1 Voronoi diagrams of the crystal structures calculated from the simulation for four different conditions- i, ii, iii, and iv. The results for $m = 0.88$ size ratio where the two particle sizes are closer (i and iii) show almost no difference in the overall hexatic order for different number ratios. Both cases (90:1 and 1:1 number ratios) show defects along the boundaries of different crystalline grains. In contrast, when the size ratio is $m = 0.80$, there is notable difference in the hexatic order (ii and iv). While the 90:1 sample remains crystalline because of the smaller number of impurities, the 1:1 sample loses its hexatic order shown by the 5- and 7-fold defects.

## 2. Simulation results for additional conditions

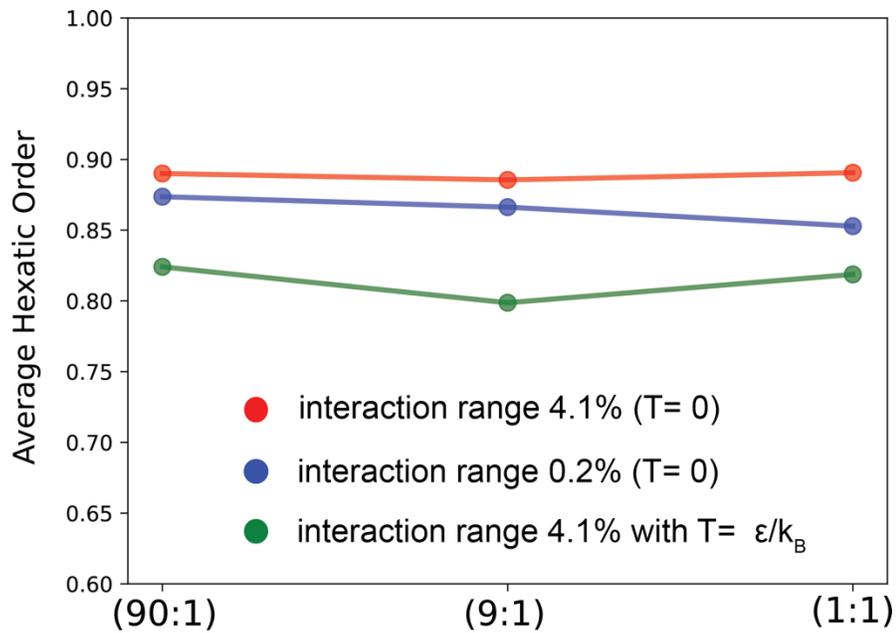

Figure S2 Average hexatic order vs number ratio plots for a fixed size ratio $m = 0.88$. Three different simulation conditions were applied: red - the range of attractive interaction $\sigma$ was 4.1% of particle diameter and the final temperature of the simulation was close to zero, blue - the range of attractive interaction $\sigma$ was 0.2% of particle diameter and the final temperature of the simulation was close to zero, green - the range of attractive interaction $\sigma$ was 4.1% of particle diameter and the final temperature of the simulation was close to the room temperature.

## 3. Difference in growth mechanism of the monodisperse and binary mixture

### a  Monodisperse

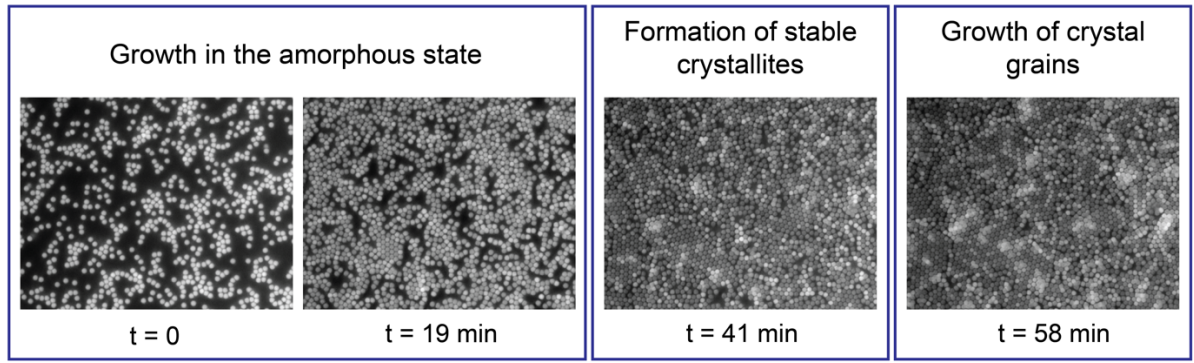

### b  Binary

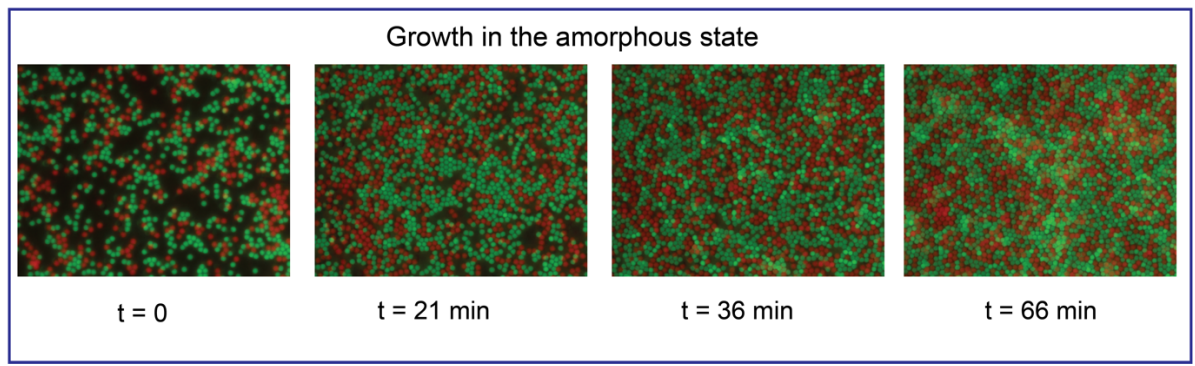

Figure S3 (a) The two-step nucleation and growth mechanism observed in the monodisperse sample. The assembly first grows in an amorphous state (t = 0 and t = 19 min), and stable crystallites start forming in the second step (t = 41 min) resulting a polycrystalline monolayer (t = 58 min). (b) The assembly grows in the amorphous state in a binary sample from the beginning (t = 0) until the complete formation of a monolayer (t = 66 min). Since stable crystallites never form, the particles remain in an arrested state with a reduced structural order in the assembly.

4. Colloidal gel formation in a binary sample for stronger interaction potential

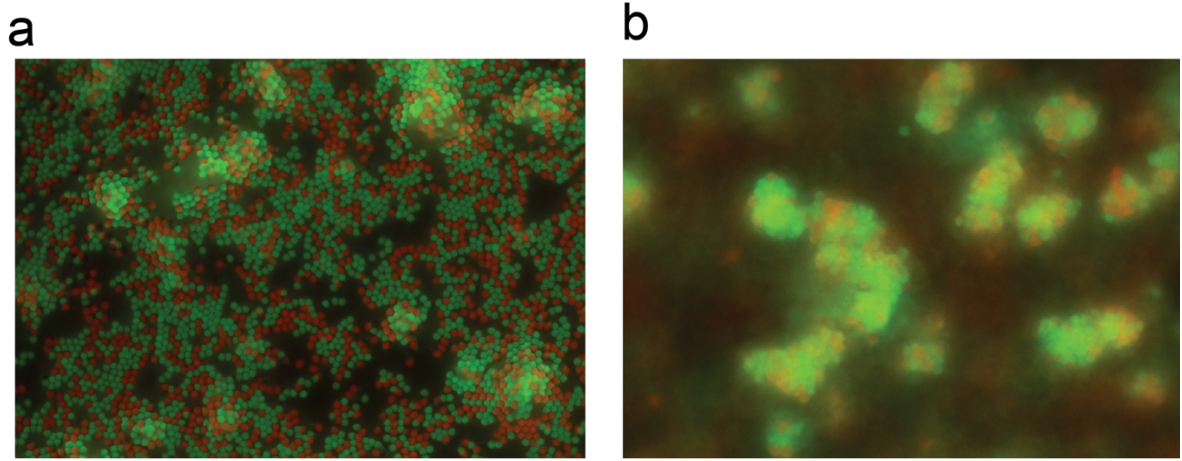

Figure S4 Formation of colloidal gels in a mixture of 700 nm and 790 nm particles and 42 mM SDS (depletant). (a) The monolayer formed at the bottom wall of the sample chamber. (b) 3D gels formed above the wall in the suspension.

## Author contributions

S.K.T.S performed the experiments, analyzed the experimental data, and prepared the manuscript. C.F. performed the Molecular Dynamics simulations. S.M. developed the computational methods used for image processing and data analysis. N.T. supervised the project and prepared the manuscript.

## Acknowledgements

We acknowledge the support from the National Science Foundation (NSF), award no. CBET – 2301692.